\documentstyle[pra,aps,amsfonts]{revtex}
\begin{document}
\draft
\title{A double-slit quantum eraser}
\author{S. P. Walborn $^{1}$, M. O. Terra Cunha$^{1,2}$, S. 
P\'adua$^{1}$,
and C. H. Monken$^{1}$}
\address{$^{1}$ Departamento de F\'{\i}sica, 
Universidade Federal de Minas Gerais, Caixa Postal 702, 
Belo Horizonte, MG 30123-970, Brazil \\
$^{2}$ Departamento de Matem\'atica, 
Universidade Federal de Minas Gerais, Caixa Postal 702, 
Belo Horizonte, MG 30123-970, Brazil} 
\date{\today}
\maketitle
\begin{abstract}
We report a quantum eraser experiment which actually uses a Young
double-slit to create interference.  The experiment can be considered
an optical analogy of an experiment proposed by Scully, Englert and
Walther(SEW){[Nature {\bf 351}, 111 (1991)]}.  One photon of an
entangled pair is incident on a Young double-slit of appropriate
dimensions to create an interference pattern in a distant detection
region.  Quarter-wave plates, oriented so that their fast axes are
orthogonal, are placed in front of each slit to serve as which-path
markers.  The quarter-wave plates mark the polarization of the
interfering photon and thus destroy the interference pattern.  To
recover interference, we measure the polarization of the other
entangled photon.  In addition, we perform the experiment under
``delayed erasure" circumstances.

\end{abstract}
\pacs{42.50.Ar, 42.25.Kb}

\section{introduction}

Wave-particle duality, a manifestation of the complementarity
principle, proposes many questions about the behavior of particles in
interferometers.  It has long been known that which-path information
and visibility of interference fringes are complementary quantities:
any distinguishability between the paths of an interferometer destroys
the quality (visibility) of the interference fringes.  The
incompatibility between which-path information and interference
effects has been quantified through inequalities by various authors
\cite{wootters79,greenberger88,mandel91,jaeger95,englert96,bjork98}. 
Originally, it was thought that the uncertainty principle was the
mechanism responsible for the absence of interference fringes due to a
which-path measurement.  The first and perhaps most famous example of
this idea is the Einstein-Bohr dialogue at the Fifth Solvay conference
in Brussels concerning Einstein's recoiling double-slit gedanken
experiment, in which the momentum transfer from incident particles to
the double-slit is measured to determine the particles'
trajectories\cite{jammer74,belinfante75}.  However, Bohr showed that
the uncertainty in the knowledge of the double-slit's initial position
was of the same order of magnitude as the space between the
interference minima and maxima: interference fringes were ``washed
out" due to the uncertainty principle\cite{jammer74}.

More recently, Scully and Dr{\"u}hl have shown that, in certain cases,
we can attribute this loss of interference not to the uncertainty
principle but to quantum entanglement between the interfering
particles and the measuring apparatus\cite{scully82}.
 
For example, disregarding internal degrees of freedom, we can
represent the state of particles exiting an interferometer by
\begin{equation}
|\Psi\rangle = \frac{1}{\sqrt{2}}(
|\psi_{1}(\bbox{\mathrm{r}})\rangle+|\psi_{2}(\bbox{\mathrm{r}})\rangle),
\label{eq:int1}
\end{equation}
where $|\psi_{1}(\bbox{\mathrm{r}})\rangle$ and 
$|\psi_{2}(\bbox{\mathrm{r}})\rangle$
represent the possibility for the particles to take path $1$ or $2$,
respectively.  The probability distribution for one particle detection
at a point $\bbox{\mathrm{r}}$ is given by
$|\langle\bbox{\mathrm{r}}|\Psi\rangle|^{2}$; the cross terms
$\langle\psi_{1}(\bbox{\mathrm{r}})|\bbox{\mathrm{r}}\rangle%
\langle\bbox{\mathrm{r}}|\psi_{2}(\bbox{\mathrm{r}})\rangle$
and
$\langle\psi_{2}(\bbox{\mathrm{r}})|\bbox{\mathrm{r}}\rangle%
\langle\bbox{\mathrm{r}}|\psi_{1}(\bbox{\mathrm{r}})\rangle$ are
responsible for interference.  The introduction of an apparatus $M$
capable of marking the path taken by a particle without disturbing
$|\psi_{1}(\bbox{\mathrm{r}})\rangle$ or
$|\psi_{2}(\bbox{\mathrm{r}})\rangle$ can be represented by the
expansion of the Hilbert space of the system in the following way:
\begin{equation} 
|\Psi\rangle = \frac{1}{\sqrt{2}}(
|\psi_{1}(\bbox{\mathrm{r}})\rangle|M_{1}\rangle+
|\psi_{2}(\bbox{\mathrm{r}})\rangle|M_{2}\rangle),
\label{eq:int2}
\end{equation}
where $|M_{j}\rangle$ is the state of the which-path marker
corresponding to the possibility of passage through the path $j$.  The
which-path marker has become entangled with the two possible particle
states.  A $100$\% effective which-path marker is prepared such that
$|M_{1}\rangle$ is orthogonal to $|M_{2}\rangle$.  In this case, a
measurement of $M$ reduces $|\Psi\rangle$ to the appropriate state for
the passage of the particle through path $1$ or $2$.  However, the
disappearance of the interference pattern is not dependent on such a
measurement.  The which-path marker's presence alone is sufficient to
make the two terms on the right-hand side of equation (\ref{eq:int2})
orthogonal and thus there will be no cross terms in
$|\langle\bbox{\mathrm{r}}|\Psi\rangle|^{2}$.  Therefore, it is enough
that the which-path information is available to destroy interference. 
Moreover, provided that $|\psi_{1}(\bbox{\mathrm{r}})\rangle$ and
$|\psi_{2}(\bbox{\mathrm{r}})\rangle$ are not significantly perturbed
by the observer, one can \emph{erase} the which-path information and
recover interference by correlating the particle detection with an
appropriate measurement on the which-path markers.  Such a measurement
is known as \emph{quantum erasure}.  In addition, if the which-path
marker is capable of storing information, the erasure can be performed
even after the detection of the particle.  The possibility of delayed
erasure generated a discussion in respect to it's legitimacy, with the
argument that it would be possible, in this way, to alter the past
\cite{mohrhoff96,mohrhoff99}.  This argument is founded on an
erroneous interpretation of the formalism of quantum mechanics
\cite{scully98,englert99}.  In recent years, there have been a number
of ideas and experiments (performed and proposed) in which which-path
information is accessible without causing severe perturbations to the
interfering particles
\cite{scully91,scully82,scully89a,scully89b,zajonc91,kwiat92,%
kwiat94,herzog95,monken96,gerry96,ou97,zheng98,hackenbroich98,durr98a,%
durr98b,kim00,kwiat98,eichmann93}.  Among the proposals, we
distinguish the ones due to Scully and Dr\"uhl \cite{scully82} and to
Scully, Englert and Walther \cite{scully91} due to their originality
and pedagogical content.

Due to their momentum, time and polarization correlation properties,
photon pairs generated by spontaneous parametric down-conversion play
an important role in the experimental demonstrations of quantum
erasure \cite{zajonc91,kwiat92,herzog95,monken96,kim00,kwiat98}. 
Although the quantum erasure phenomenon is present in all reported
experiments, only one \cite{kim00} can be considered an optical analog
of the original proposal of Scully and Dr\"uhl \cite{scully82}.  In
this paper we report a quantum eraser experiment which actually uses a
Young double-slit to create interference.  The experiment is analyzed
in connection with the proposal of Scully, Englert and Walther (SEW)
\cite{scully91,scully97}.  To the authors' knowledge, this is the
first demonstration of a quantum eraser in which interference is
obtained from the passage of the particles through a real double-slit.

In section II we give a brief summary of the SEW quantum eraser.  The
theory behind our quantum eraser is presented in section III. The
experimental setup and results are presented in sections IV and V,
respectively.

\section{THE SEW QUANTUM ERASER}

The experiment reported here is inspired by the proposal of Scully,
Englert and Walther \cite{scully91}, which can be summarized as
follows: A beam of Rydberg atoms in an excited state is incident on a
double-slit small enough to form a Young interference pattern on a
distant screen.  In front of each slit is placed a which-path marker,
which consists of a micromaser cavity of appropriate length such that
the emission probability for an atom traversing the cavity is $1$. 
Then, the presence of a photon in either cavity marks the passage of
an atom through the corresponding slit and thus destroys the
interference pattern, because which-path information is now available. 
The perturbation to the spatial part of the wave function of the atoms
due to the cavities is ignorable \cite{scully91,scully98,scully97}.  A
measurement that projects the state of the cavities onto a symmetric
(antisymmetric) combination of $|0\rangle$ (no photon present) and
$|1\rangle$ (one photon present) performs the erasure, and an 
interference pattern is recovered in correlated detection.

\section{AN OPTICAL BELL-STATE QUANTUM ERASER}

Consider the following experimental setup: A linearly polarized beam
of photons is incident on a double-slit.  If the double slit is of
appropriate dimensions, the probability distribution for a one-photon
detection at a distant screen is given by a Young interference
pattern.  Suppose that in front of each slit we place a quarter-wave
plate, with the fast axis at an angle of $45^{\circ}$ (or
$-45^{\circ}$) with respect to the photon polarization direction. 
Upon traversing either one of the waveplates, the photon becomes
circularly polarized, and acquires a well-defined angular momentum
\cite{beth36}.  Supposing that the waveplate is free to rotate, it
should acquire an angular momentum opposite to that of the photon, and
rotate right or left, depending on the chirality of the photon.  If we
treat each waveplate as a quantum rotor, we can say that the photon
induced a transition with $\Delta\ell = \pm 1$.  Since the waveplates
don't significantly modify the propagation of the beam, we have, in
principle, a which-path marker with necessary characteristics for a
quantum eraser.  However, this scheme is far from being practical. 
Besides the difficulty to set the waveplates free to rotate, the
separation between the energy levels of a rotor with the mass and
dimensions of a waveplate is of the order $10^{-40}$ eV. In addition,
decoherence effects may make it impossible to use macroscopic quantum
rotors to mark the path of a photon.  This idea is similar to the
``haunted measurement" of Greenberger and Ya'sin \cite{greenberger89}.

By enlarging the system, however, it is possible to create an adequate
which-path detector.  Let the beam of photons incident on the
double-slit be entangled with a second beam freely propagating in
another direction, so as to define a Bell state
\begin{equation}
|\Psi\rangle =
\frac{1}{\sqrt{2}}(|x\rangle_{s}|y\rangle_{p}+|y\rangle_{s}|x\rangle_{p}),
\label{eq:state1}
\end{equation}
where the indices $s$ and $p$ indicate the two beams, and $x$ and $y$
represent orthogonal linear polarizations.  If beam $s$ is
incident on the double-slit (without waveplates), state
(\ref{eq:state1}) is transformed to
\begin{equation}
|\Psi\rangle=\frac{1}{\sqrt{2}}(|\psi_{1}\rangle + |\psi_{2}\rangle),
\label{eq:placa1}
\end{equation}
where
\begin{equation} 
|\psi_{1}\rangle =
\frac{1}{\sqrt{2}}(|x\rangle_{s1}|y\rangle_{p}+|y\rangle_{s1}|x\rangle_{p}),
\end{equation}
\begin{equation} 
|\psi_{2}\rangle =
\frac{1}{\sqrt{2}}(|x\rangle_{s2}|y\rangle_{p}+|y\rangle_{s2}|x\rangle_{p}).
\end{equation}
The indices $s1$ and $s2$ refer to beams generated by slit $1$ and
slit $2$, respectively.  The probability distribution for one-photon
detection on a screen placed in the far field region of the 
overlapping beams $s1$ $s2$ will show the usual interference:
\begin{equation}
P_{s}(\delta)\propto 1+\cos{\delta},
\label{eq:prob1}
\end{equation} 
where $\delta$ is the phase difference between the paths: slit
$1 \rightarrow$ detector and slit $2 \rightarrow$ detector.  
Introducing the $\lambda/4$ plates one in front of each slit with the
fast axes at angles $\theta_{1} = 45^{\circ}$  and
$\theta_{2}=-45^{\circ}$, with the $x$ direction,  
states $|\psi_{1}\rangle$ and $|\psi_{2}\rangle$ are transformed to
\begin{equation} 
|\psi_{1}\rangle =
\frac{1}{\sqrt{2}}(|L\rangle_{s1}|y\rangle_{p}+i|R\rangle_{s1}|x\rangle_{p}),
\label{eq:placa2}
\end{equation}
\begin{equation} 
|\psi_{2}\rangle =
\frac{1}{\sqrt{2}}(|R\rangle_{s2}|y\rangle_{p}-i|L\rangle_{s2}|x\rangle_{p}),         
\label{eq:placa3}
\end{equation}
where $R$ and $L$ represent right and left circular polarizations. 
Since $|\psi_{1}\rangle$ and $|\psi_{2}\rangle$ have orthogonal
polarizations, there is no possibility of interference.  In order to
recover interference, let us project the state of the system
over the symmetric and antisymmetric states of the which-path
detector.  This is equivalent to transforming $|\psi_{1}\rangle$ and
$|\psi_{2}\rangle$ in a way that expresses them as symmetric and
antisymmetric combinations of polarizations, for example,
\begin{equation}
|x\rangle =
\frac{1}{\sqrt{2}}(|+\rangle + |-\rangle),
\end{equation}
\begin{equation}
|y\rangle = \frac{1}{\sqrt{2}}(|+\rangle - |-\rangle),
\end{equation}
\begin{equation}
|R\rangle =\frac{1-i}{2}(|+\rangle + i |-\rangle),
\end{equation}
\begin{equation} 
|L\rangle = \frac{1-i}{2}(i |+\rangle + |-\rangle),
\end{equation} 
where ``$+$'' and ``$-$'' represent
polarizations $+45^{\circ}$ and $-45^{\circ}$ with respect to $x$. 
Rewriting the complete state $|\Psi\rangle$, we have 
\begin{equation}
|\Psi\rangle =
\frac{1}{2}\left[(|+\rangle_{s1}-i|+\rangle_{s2})|+\rangle_{p}+%
i(|-\rangle_{s1}+i|-\rangle_{s2})|-\rangle_{p}\right]
\label{eq:psisym}
\end{equation}
According to the above expression, we can recover interference
projecting the state of photon $p$ over $|+\rangle_{p}$ or
$|-\rangle_{p}$.  Experimentally, this can be done by placing a
polarizer in the path of beam $p$ and orientating it at +45$^{\circ}$
to select $|+\rangle_{p}$ or at -45$^{\circ}$ to select
$|-\rangle_{p}$.  The interference pattern is recovered through the
coincidence detection of photons $s$ and $p$.  Notice that the fringes
obtained in the two cases are out of phase.  They are commonly called
\emph{fringes} and \emph{anti-fringes}.

\subsection{Obtaining which-path information}
Which-path information can be obtained by considering the polarization
of both photons $s$ and $p$.  The process of obtaining information can
be separated into two schemes: Detecting $p$ before $s$, or detecting
$s$ before $p$, which we refer to as {\emph{delayed erasure}}.  This
can be done by changing the relative lengths of beams $s$ and $p$.  We
will assume that one photon is detected much earlier than the arrival
of the other photon at the measuring devices.  Let us consider the
first possibility.  If photon $p$ is detected with polarization $x$
(say), then we know that photon $s$ has polarization $y$ before
hitting the $\lambda/4$ plates and the double slit.  By looking at
equations (\ref{eq:placa1}), (\ref{eq:placa2}) and (\ref{eq:placa3}))
it is clear that detection of photon $s$ (after the double slit) with
polarization $R$ is compatible only with the passage of $s$ through
slit $1$ and polarization $L$ is compatible only with the passage of
$s$ through slit $2$.  This can be verified experimentally.  In usual
quantum mechanics language, detection of photon $p$ before photon $s$
has prepared photon $s$ in a certain state.

\subsection{Delayed erasure}

The possibility of obtaining which-path information after the
detection of photon $s$ leads to delayed choice \cite{wheeler}. 
Delayed choice creates situations in which it is important to have a
clear notion of the physical significance of quantum mechanics.  A
good discussion can be found in references
\cite{mohrhoff96,mohrhoff99,scully98,englert99}.  In as much as our
quantum eraser does not allow the experimenter to \emph{choose} to
observe which-path information or an interference pattern after the
detection of photon $s$, it does allow for the detection of photon $s$
before photon $p$, a situation to which we refer to as \emph{delayed
erasure}.  The question is: ``Does the order of detection of the two
photons affect the experimental results?"

\section{EXPERIMENTAL SETUP AND PROCEDURE}

For certain propagation directions, type II spontaneous parametric
down conversion (SPDC) in a nonlinear crystal creates the state
\begin{equation}
|\Psi\rangle=\frac{1}{\sqrt{2}} ( |o\rangle_{s}|e\rangle_{p} + e^{i
\phi}|e\rangle_{s}|o\rangle_{p} ),
\end{equation}
where $o$ and $e$ refer to ordinary and extraordinary polarizations. 
$\phi$ is a relative phase shift due to the crystal birrefringence.  If
$\phi = 0$ or $\pi$ we have the Bell states $|\Psi^+\rangle$ and 
$|\Psi^-\rangle$,
respectively.  

Using this state in the interferometer described in the previous 
section, the probability of detecting photons in coincidence is
proportional to
\begin{equation}
\frac{1}{2}+\left[ \frac{1}{2} - \sin^2(\theta+\alpha)\cos^2%
\frac{\phi}{2} - \sin^2(\theta-\alpha)\sin^2\frac{\phi}{2} \right]\sin{\delta}, 
\end{equation}
where $\delta$ is defined right after expression (\ref{eq:prob1}),
$\theta$ is the smallest angle between the fast (slow) axis of the
quarter-wave plates and the $o$ axis and $\alpha$ is the angle of the
polarizer in path $p$, with respect to the $o$ axis.

The experimental setup is shown in FIG \ref{fig1}.  An Argon laser
(351.1 nm at $\sim 200$ mW) is used to pump a 1mm long BBO
($\beta$-BaB$_2$O$_4$) crystal, generating $702.2$ nm entangled
photons by spontaneous parametric down-conversion.  The BBO crystal
is cut for type-II phase matching.  The pump beam is focused onto the
crystal plane using a 1 m focal length lens to the increase transverse
coherence length at the double-slit.  The width of the pump beam at
the focus is approximately $0.5$mm\cite{ribeiro94}.  The orthogonally
polarized entangled photons leave the BBO crystal each at an angle of
$\sim 3^\circ$ with the pump beam.  In the path of photon $p$ a
polarizer cube (POL1) can be inserted in order to perform the
quantum erasure.  The double-slit and quarter-wave plates are placed
in path $s$, 42 cm from the BBO crystal.  Detectors $D_s$ and $D_p$
are located 125 cm and 98 cm from the BBO crystal, respectively.  QWP1
and QWP2 are quarter-wave plates with fast axes at an angle of
45$^\circ$.  The circular quarter-wave plates were sanded
(tangentially) so as to fit together in front of the double-slit.  The
openings of the double-slit are $200 \mu$m wide and separated by a
distance of $200 \mu$m.  The detectors are EG\&~G SPCM 200
photodetectors, equipped with interference filters (bandwidth 1 nm)
and $300 \mu$m $\times$ $5$ mm rectangular collection slits.  A
stepping motor is used to scan detector $D_s$.

The delayed erasure setup is similar, with two changes: ($i$)
detector $D_p$ and POL1 were placed at a new distance of 2 meters
from the BBO crystal and ($ii$) the collection iris on detector $D_p$
has dimensions $600 \mu$m $\times$ $5$ mm.
 
\section{EXPERIMENTAL RESULTS}

Before the quantum eraser experiment was performed, Bell's inequality
tests were performed to verify that entangled states were being
detected \cite{kwiat95}.  Figure \ref{fig2} shows the standard Young
interference pattern obtained with the double-slit placed in the
path of photon $s$, without quarter-wave plates QWP1 and QWP2, and
with POL1 absent from detector $D_p$.  Next, the path of photon $s$
was marked by placing the quarter-wave plates QWP1 and QWP2 in front
of the double-slit.  Figure \ref{fig3} shows the absence of
interference due to the quarter-wave plates.  Nearly all interference
present in figure \ref{fig2} was destroyed.  The residual interference
present is due to a small error in aligning the quarter-wave plates. 
The which-path information was \emph{erased} and interference
recovered by placing the linear polarizer POL1 in front of detector
$D_p$.  To recover interference, the polarization angle of POL1
($\alpha$) was set to $\theta$, the angle of the fast axis of
quarter-wave plate QWP1.  Interference fringes were obtained as shown
in figure \ref{fig4}.  The detection time was doubled in order to
compensate for the decrease in coincidence counts due to POL1.  In
figure \ref{fig5}, POL1 was set to $\theta+\frac{\pi}{2}$, the angle
of the fast axis of QWP2, which produced a pattern of interference
anti-fringes.  The averaged sum of these two interference patterns
gives a pattern roughly equal to that of figure \ref{fig3}.

The same experimental procedure was used to produce figures \ref{fig6}
- \ref{fig9} for the delayed erasure situation.  The experimental
results are comparable to the case in which photon $p$ is detected
before photon $s$.  We use the term ``delayed choice" loosely, in that
in our experiment there is no ``choice" available to the observer in
the time period after the detection of photons $s$ and before the
detection of photon $p$.  We simply wish to show that the order of
detection is not important, in concordance with the
literature\cite{scully98,englert99}.

\section{CONCLUSION}

We have presented a quantum eraser which uses a Young double-slit to
create interference.  The quarter-wave plates in our experiment served
as the which-path markers to destroy interference.  We recovered
interference using the entanglement of photons $s$ and $p$.  Our
quantum eraser is very similar to the that of Scully, Englert and
Walther\cite{scully91}.  We have shown that interference can be
destroyed, by marking the path of the interfering photon, and
recovered, by making an appropriate measurement on the other entangled
photon.  We have also investigated this experiment under the
conditions of delayed erasure, in which the interfering photon $s$ is
detected before photon $p$.  In as much as our experiment did not
allow for the observer to \emph{choose} the polarization angle in the
time period after photon $s$ was detected and before detection of $p$,
our results show that a \emph{collapse} of the wave function due to
detection of photon $s$ does not prohibit one from observing the
expected results.  Our experimental data agrees with the proposal of
Scully, Englert and Walther, that quantum erasure can be performed
after the interfering particle has been detected \cite{scully91}.

\section{ACKNOWLEDGMENT}
The authors acknowledge financial support from the Brazilian 
agencies CNPq, CAPES, FINEP, PRONEX, and FAPEMIG. 


\begin{thebibliography}{99}

\bibitem{wootters79}
W.~K. Wootters and W.~H. Zurek, Phys.  Rev.  D {\bf 19}, 473 (1979).

\bibitem{greenberger88}
D.~M. Greenberger and A. Yasin, Phys.  Lett.  A {\bf 128}, 391 (1988).

\bibitem{mandel91}
L. Mandel, Opt.  Lett.  {\bf 16}, 1882 (1991).

\bibitem{jaeger95}
G. Jaeger, A. Shimony, and L. Vaidman, Phys.  Rev.  A {\bf 51}, 54
(1995).

\bibitem{englert96}
B.~G. Englert, Phys.  Rev.  Lett.  {\bf 77}, 2154 (1996).

\bibitem{bjork98}
G. Bj\"ork and A. Karlsson, Phys.  Rev.  A {\bf 58}, 3477 (1998).

\bibitem{jammer74}
M. Jammer, {\em The Philosophy of Quantum Mechanics} (Wiley, New York,
1974).

\bibitem{belinfante75}
F. Belinfante, {\em Measurement and Time Reversal in Objective Quantum
Theory} (Pergammon, New York, 1975).

\bibitem{scully82}
M.~O. Scully and K. Dr\"uhl, Phys.  Rev.  A {\bf 25}, 2208 (1982).

\bibitem{scully91}
M.~O. Scully, B.~G. Englert, and H. Walther, Nature {\bf 351}, 111
(1991).

\bibitem{mohrhoff96}
U. Mohrhoff, Am.  J. Phys.  {\bf 64}, 1468 (1996).

\bibitem{mohrhoff99}
U. Mohrhoff, Am.  J. Phys.  {\bf 67}, 330 (1999).

\bibitem{scully98}
sM.~O. Scully and H. Walther, Found.  Phys.  {\bf 28}, 5229 (1998).

\bibitem{englert99}
B.~G. Englert, M.~O. Scully, and H. Walther, Am.  J. Phys.  {\bf 67},
325 (1999).

\bibitem{scully89a}
M.~O. Scully and H. Walther, Phys.  Rev.  A {\bf 39}, 5229 (1989).

\bibitem{scully89b}
M.~O. Scully, B.~G. Englert, and J. Schwinger, Phys.  Rev.  A {\bf
40}, 1775 (1989).

\bibitem{zajonc91}
A.~G. Zajonc, L. Wang, X. Ou, and L. Mandel, Nature {\bf 353}, 507
(1991).

\bibitem{kwiat92}
P.~G. Kwiat, A. Steinberg, and R. Chiao, Phys.  Rev.A {\bf 45}, 7729
(1992).

\bibitem{kwiat94}
P.~G. Kwiat, A. Steinberg, and R. Chiao, Phys.  Rev.A {\bf 49}, 7729
(1994).

\bibitem{herzog95}
T.~J. Herzog, P.~G. Kwiat, H. Weinfurter, and A. Zeilinger, Phys. 
Rev.  Lett.  {\bf 75}, 3034 (1995).

\bibitem{monken96}
C.~H. Monken, D. Branning, and L. Mandel, in {\em CQO7 - Proceedings
of the 7$\underline{th}$ Rochester Conference in Coherence and Quantum
Optics}, edited by L. Mandel and J. Eberly (Plenum Press, New York,
1996), p.\ 701.

\bibitem{gerry96}
C.~C. Gerry, Phys.  Rev.  A {\bf 53}, 1179 (1996).

\bibitem{ou97}
Z.~Y. Ou, Phys.  Lett.  A {\bf 226}, 323 (1997).

\bibitem{zheng98}
S.-B. Zheng and G.-C. Guo, Physica A {\bf 251}, 507 (1998).

\bibitem{hackenbroich98}
G. Hackenbroich, B. Rosenow, and H.~A. Weidenm\"uller, Europhys. 
Lett.  {\bf 44}, 693 (1998).

\bibitem{durr98a}
S. D\"urr, T. Nonn, and G. Rempe, Nature {\bf 395}, 33 (1998).

\bibitem{durr98b}
S. D\"urr, T. Nonn, and G. Rempe, Phys.  Rev.  Lett.  {\bf 81}, 5705
(1998).

\bibitem{kim00}
Y. Kim, S. Kulik, Y. Shih, and M. Scully, Phys.  Rev.  Lett.  {\bf
84}, 1 (2000).

\bibitem{kwiat98}
P. Kwiat, P. Schwindt, and B.~G. Englert, in {\em Workshop on
Mysteries, Puzzles and Paradoxes in Quantum Mechanics}, edited by R.
Bonifacio (AIP, Gargnano, Italy, 1998).

\bibitem{eichmann93}
U. Eichmann {\it et~al.}, Phys.  Rev.  Lett.  {\bf 70}, 2359 (1993).

\bibitem{scully97}
M.~O. Scully and M.~S. Zubairy, {\em Quantum Optics} (Cambridge
University Press, Cambridge, 1997).
\bibitem{beth36}
R.~A. Beth, Phys.  Rev.  {\bf 50}, 115, 1936.

\bibitem{greenberger89}
D.~M. Greenberger and A. Yasin, Found.  Phys.  {\bf 19}, 679 (1989).

\bibitem{wheeler}
J.~A. Wheeler, in {\em Mathematical Foundations of Quantum Theory},
J.~A. Wheeler and W.~H. Zurek Eds., 183, Princeton Univ.  Press,
Princeton, N.J.
\bibitem{ribeiro94}
P.~S. Ribeiro, C. Monken, and G. Barbosa, Applied Optics {\bf 33}, 352
(1994).

\bibitem{kwiat95}
P.~G. Kwiat {\it et~al.}, Phys.  Rev.  Lett.  {\bf 75}, 4337 (1995).

\end{thebibliography}


\begin{figure}
\caption{Experimental setup for the Bell-state quantum eraser.  
QWP1 and QWP2 are quarter-wave plates aligned in front of the double 
slit with fast
axes perpendicular.  POL1
is a linear polarizer.}
\label{fig1}
\end{figure}
\begin{figure}
\caption{Coincidence counts vs. detector D$_s$ position with QWP1 and QWP2
removed.  An 
 Interference pattern due to the double-slit is observed.}
\label{fig2}
\end{figure}
\begin{figure}
\caption{Coincidence counts when QWP1 and QWP2 are placed in front of 
the double-slit. Interference has been destroyed.}
\label{fig3}
\end{figure}
\begin{figure}
\caption{Coincidence counts when QPW1, QWP2 and POL1 are in place.  
POL1 was set to $\theta$, the angle of the fast axis of QWP1.
Interference has been restored in the \emph{fringe} pattern.}
\label{fig4}
\end{figure}
\begin{figure}
\caption{Coincidence counts when QPW1, QWP2 and POL1 are in place.  
POL1 was set to $\theta + \frac{\pi}{2}$, the angle of the
fast axis of QWP2. Interference has been restored in the  
\emph{antifringe} pattern.}
\label{fig5}
\end{figure}
\begin{figure}
\caption{Coincidence counts in the delayed erasure setup.  
QWP1, QWP2 and POL1 are absent. A standard Young
interference pattern is observed.}
\label{fig6}
\end{figure}
\begin{figure}
\caption{Coincidence counts in the delayed erasure setup with  QWP1 and 
QWP2 in place in front of the double-slit. No interference is observed.}
\label{fig7}
\end{figure}
\begin{figure}
\caption{Coincidence counts in the delayed erasure setup when QPW1, 
QWP2 and 
POL1 are in place.
POL1 was set to $\theta$, the angle of the
fast axis of QWP1. Interference has been restored in the  
\emph{fringe} pattern.}
\label{fig8}
\end{figure}
\begin{figure}
\caption{Coincidence counts in the delayed erasure setup when QPW1, 
QWP2 and POL1 are in place.
POL1 was set to $\theta + \frac{\pi}{2}$, the angle of the
fast axis of QWP2. Interference has been restored in the  
\emph{antifringe} pattern.}
\label{fig9}
\end{figure}

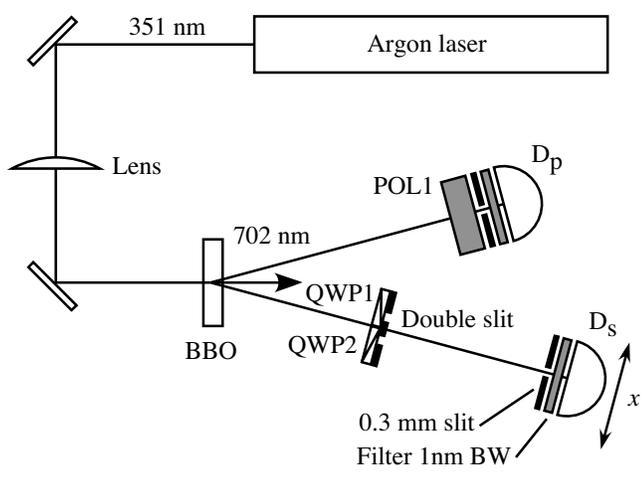
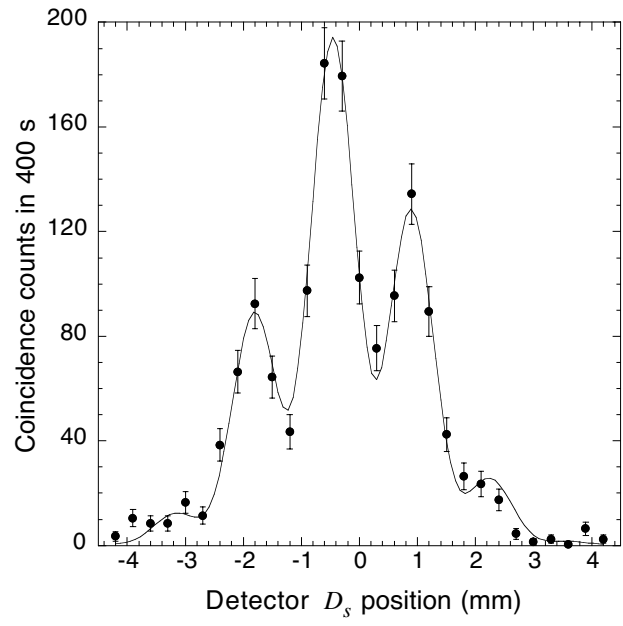
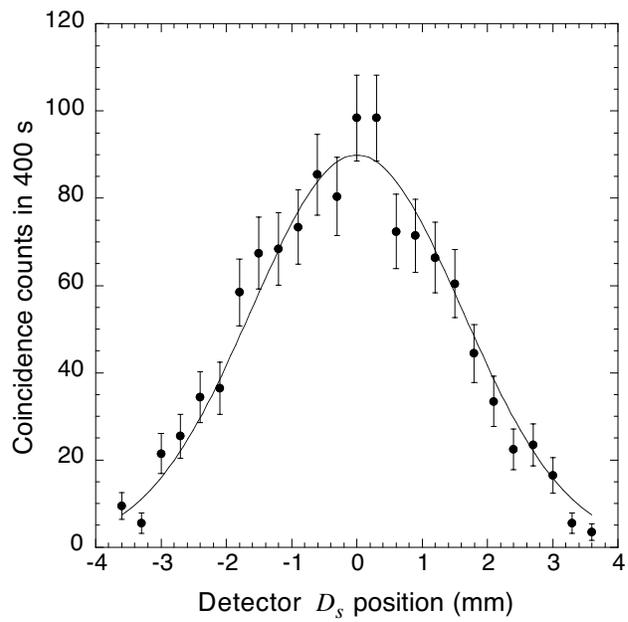
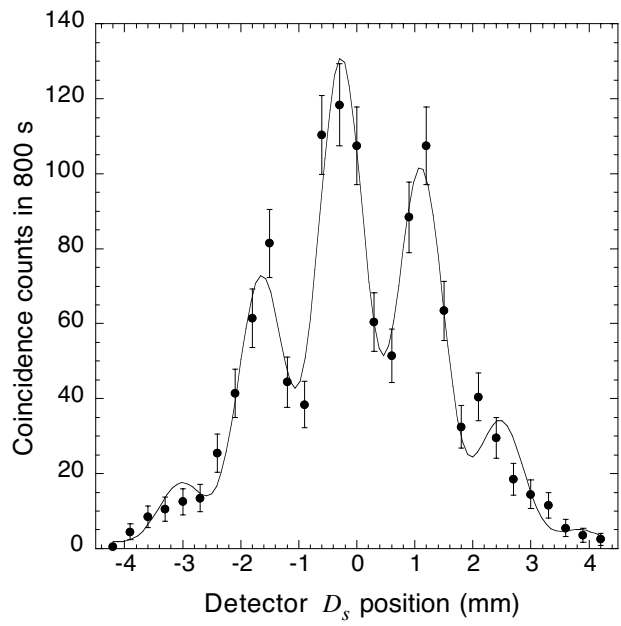
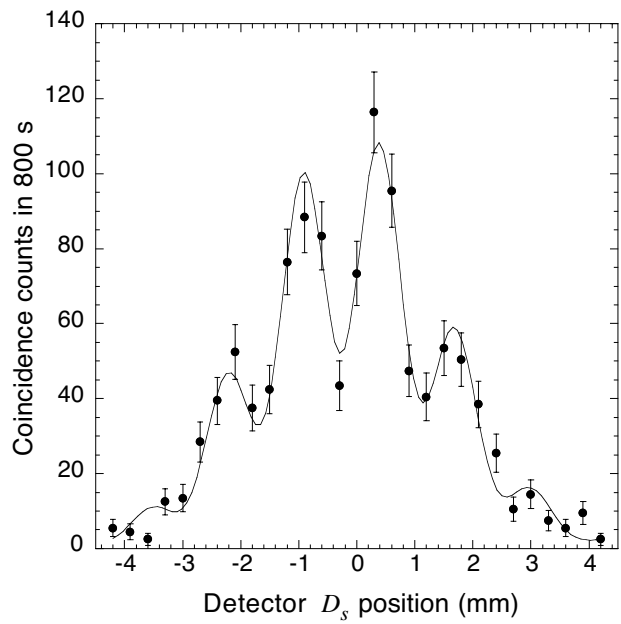
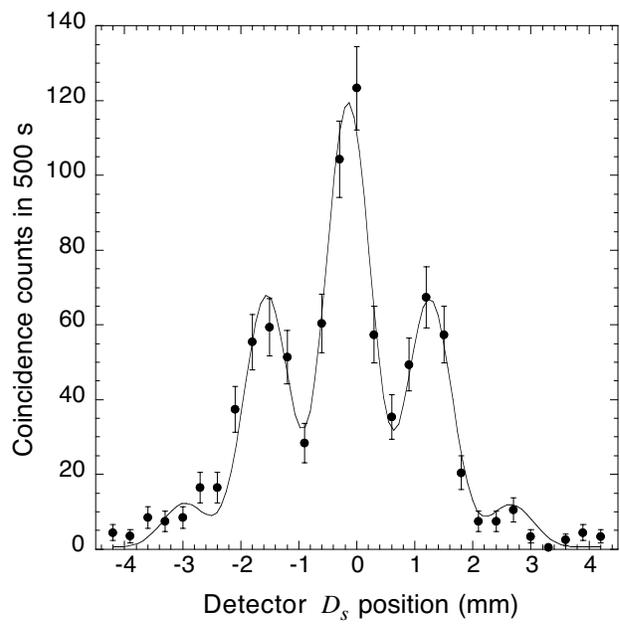
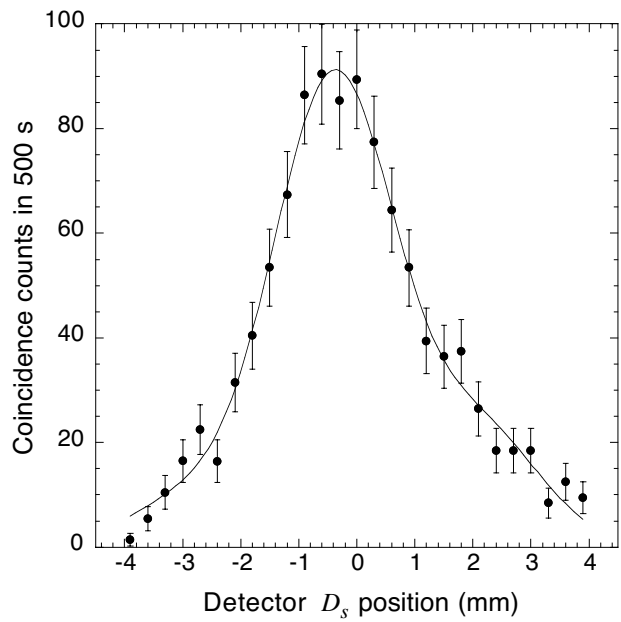
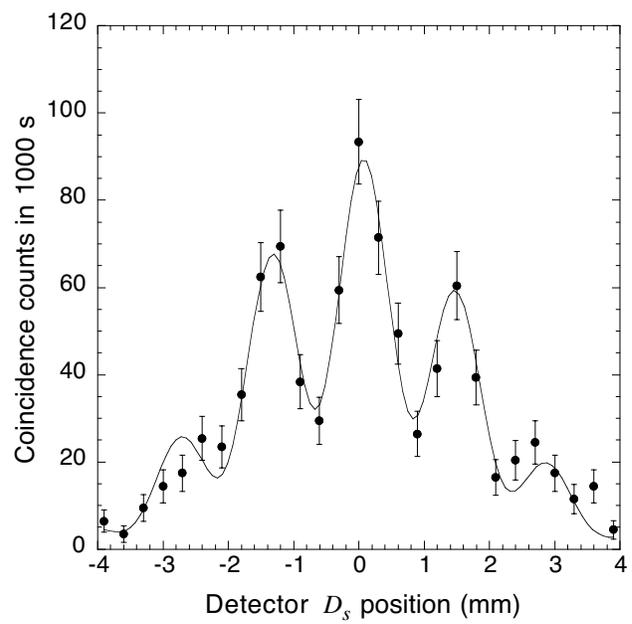
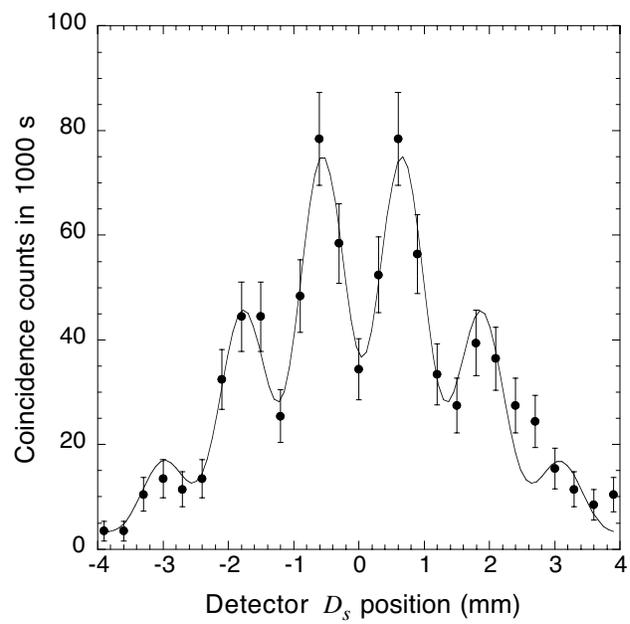
\end{document}